\documentclass{article}

\usepackage{graphicx}

\begin{document}

\def\thefootnote{\fnsymbol{footnote}}
\thispagestyle{empty}

\hbox to \hsize{%
{\large\sffamily\bfseries University of Wisconsin - Madison}\hfill
\vtop{%
\hbox{\sffamily\bfseries MADPH-04-1407}\hbox{\sffamily December 2004}}}
\baselineskip 18pt

\vskip.75in

\begin{center}
{\Large\bf High-Energy Neutrino Astronomy}\footnote{Talk presented at the Nobel Symposium 129: Neutrino Physics, Enk\"oping, Sweden, August 2004}\\[.2in]
{\large Francis Halzen}\\[.1in]
{\it Department of Physics, University of Wisconsin, Madison, WI,  53706, USA}
\end{center}

\vskip.75in

\begin{abstract}
Kilometer-scale neutrino detectors such as IceCube are discovery
 instruments covering nuclear and particle physics, cosmology and astronomy. Examples of their multidisciplinary missions include the search for the particle nature of dark matter and for additional small dimensions of space. In the end, their conceptual design is very much anchored to the observational fact that Nature accelerates protons and photons to energies in excess of $10^{20}$ and $10^{13}$\,eV, respectively. The cosmic ray connection sets the scale of cosmic neutrino fluxes. In this context, we discuss the first results of the completed AMANDA detector and the reach of its extension, IceCube. Similar experiments are under construction in the Mediterranean. Neutrino astronomy is also expanding in new directions with efforts to detect air showers, acoustic and radio signals initiated by neutrinos with energies similar to those of the highest energy cosmic rays.
\end{abstract}

\newpage

\section{Neutrinos Associated with the Highest Energy Cosmic Rays}

The flux of cosmic rays is summarized in Fig.\,1a,b\cite{gaisseramsterdam}. The  energy spectrum follows a broken power law. The two power laws are separated by a feature dubbed the ``knee''; see Fig.\,1a. Circumstantial evidence exists that cosmic rays, up to EeV energy, originate in galactic supernova remnants. Any association with our galaxy disappears however in the vicinity of a second feature in the spectrum referred to as the ``ankle''. Above the ankle, the gyroradius of a proton in the galactic magnetic field exceeds the size of the galaxy and it is generally assumed that we are  witnessing the onset of an extragalactic component in the spectrum that extends to energies beyond 100\,EeV. Experiments indicate that the highest energy cosmic rays are predominantly protons or, possibly, nuclei. Above a threshold of 50 EeV these protons interact with cosmic microwave photons and lose their energy to pions before reaching our detectors. This is the Greissen-Zatsepin-Kuzmin cutoff that limits the sources to our supercluster of galaxies.

\begin{figure}[h]
\centering\leavevmode
\includegraphics[width=6in]{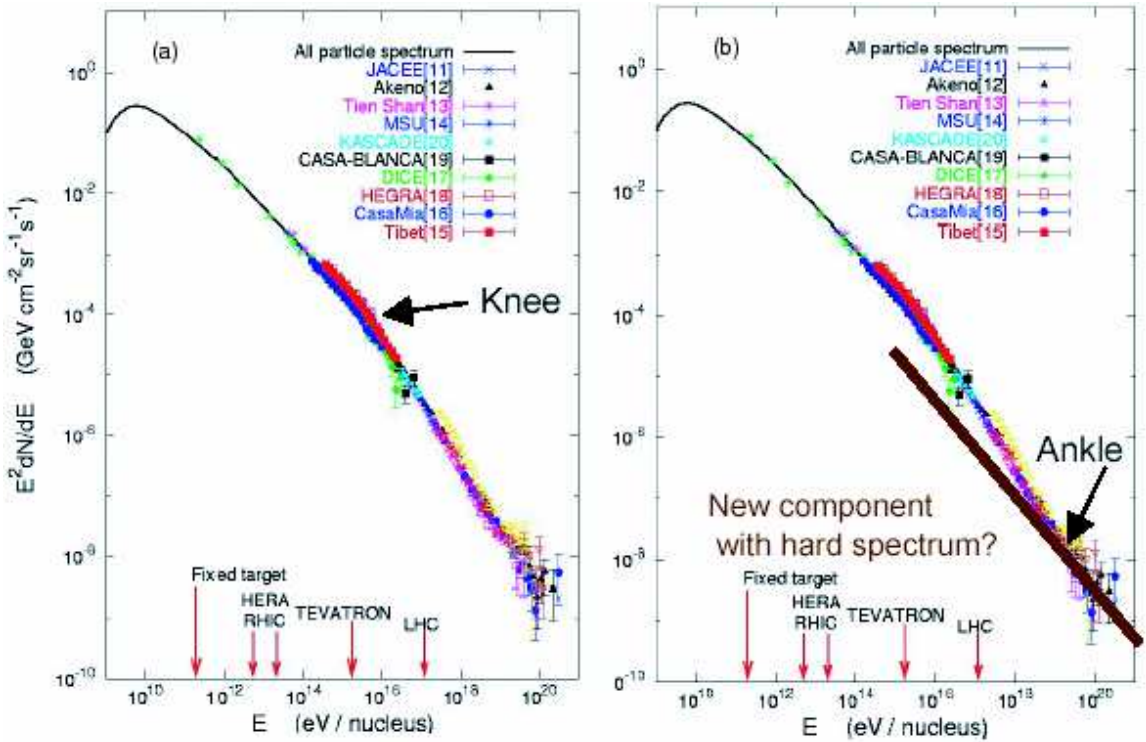}
\caption{At the energies of interest here, the cosmic ray spectrum consists of a sequence of 3 power laws. The first two are separated by the ``knee'' (left panel), the second and third by the ``ankle''. There is evidence that the cosmic rays beyond the ankle are a new population of particles produced in extragalactic sources; see right panel.}
\end{figure}

Models for the origin of the highest energy cosmic rays fall into two categories, top-down and bottom-up. In top-down models it is assumed that the cosmic rays are the decay products of cosmological remnants with Grand Unified energy scale $M_{GUT} \sim 10^{24}\rm\,eV$. These models predict neutrino fluxes most likely within reach of first-generation telescopes such as AMANDA, and certainly detectable by future kilometer-scale neutrino observatories\cite{PR}.

In bottom-up scenarios it is assumed that cosmic rays originate in cosmic accelerators. Accelerating particles to TeV energy and above requires massive bulk flows of relativistic charged particles. These are likely to originate from the exceptional gravitational forces  in the vicinity of black holes. Examples include the dense cores of exploding stars, inflows onto supermassive black holes at the centers of active galaxies and annihilating black holes or neutron stars. Before leaving the source, accelerated particles pass through intense radiation fields or dense clouds of gas surrounding the black hole. This results in interactions producing pions decaying into secondary photons and neutrinos that accompany the primary cosmic ray beam as illustrated in Fig.\,2.

\begin{figure}[h]
\centering\leavevmode
\includegraphics[width=4.25in]{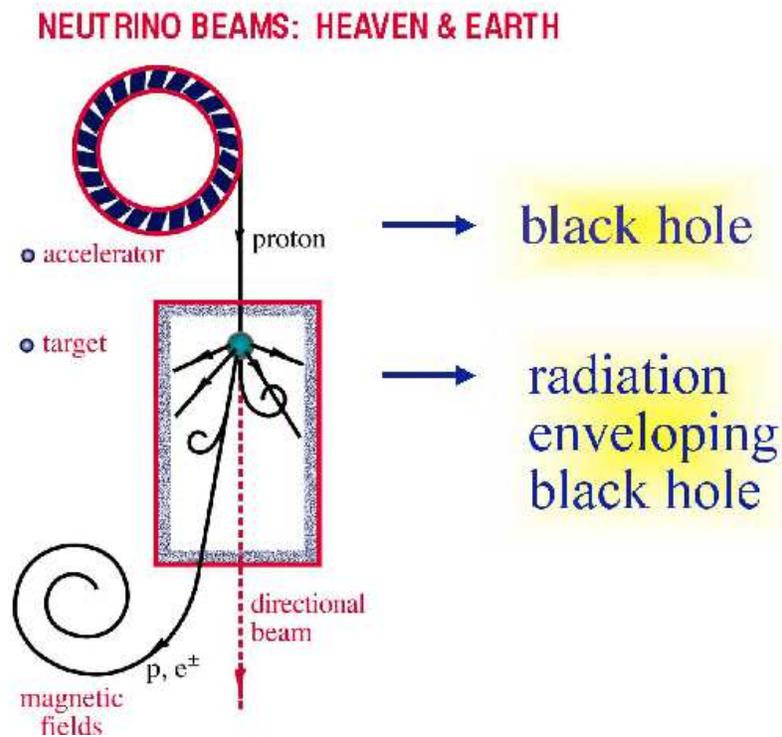}
\caption{Cosmic beam dump: sketch of cosmic ray accelerator producing photons and neutrinos.}
\end{figure}

How many neutrinos are produced in association with the cosmic ray beam? The answer to this question, among many others\cite{PR},  provides the rationale for building kilometer-scale neutrino detectors. We first consider a neutrino beam produced at an accelerator laboratory; see Fig.\,2. Here the target absorbs all parent protons as well as the secondary electromagnetic and hadronic showers. Only neutrinos exit the dump. If nature constructed such a ``hidden source'' in the heavens, conventional astronomy will not reveal it. It cannot be the source of the cosmic rays, however, because in this case the dump must be transparent to protons.

A more generic ``transparent'' source can be imagined as follows: protons are accelerated in a region of high magnetic fields where they interact with photons via the processes 
%
$p + \gamma \rightarrow \Delta \rightarrow \pi^0 + p$,
$p + \gamma \rightarrow \Delta \rightarrow \pi^+ + n$.
%
While the secondary protons may remain trapped in the acceleration region, equal numbers of neutrons, neutral and charged pions escape. The energy escaping the source is therefore equally distributed between cosmic rays, gamma rays and neutrinos produced by the decay of neutrons and neutral and charged pions, respectively. The neutrino flux from a generic transparent cosmic ray source is often referred to as the Waxman-Bahcall flux\cite{wb1}. It is easy to calculate and the derivation is revealing.

Figure 1b shows a fit to the observed spectrum above the ``ankle'' that can be used to derive the total energy in extragalactic cosmic rays. The energy content of this component is $\sim 3 \times 10^{-19}\rm\,erg\ cm^{-3}$, assuming an $E^{-2}$ energy spectrum with a GZK cutoff. The power required for a population of sources to generate this energy density over the Hubble time of $10^{10}$\,years is $\sim 3 \times 10^{37}\rm\,erg\ s^{-1}$ per (Mpc)$^3$ or, as often quoted in the literature,
 $\sim 5\times10^{44}\rm\,TeV$ per year per (Mpc)$^3$. This works out to\cite{TKG}
\begin{itemize}
\item $\sim 3 \times 10^{39}\rm\,erg\ s^{-1}$ per galaxy,
\item $\sim 3 \times 10^{42}\rm\,erg\ s^{-1}$ per cluster of galaxies,
\item $\sim 2 \times 10^{44}\rm\,erg\ s^{-1}$ per active galaxy, or
\item $\sim 2 \times 10^{52}$\,erg per cosmological gamma ray burst.
\end{itemize}
The coincidence between these numbers and the observed output in electromagnetic energy of these sources explains why they have emerged as the leading candidates for the cosmic ray accelerators. The coincidence is consistent with the relationship between cosmic rays and photons built into the ``transparent'' source. In the photoproduction processes roughly equal energy goes into the secondary neutrons, neutral and charged pions whose energy ends up in cosmic rays, gamma rays and neutrinos, respectively.

We therefore assume that the same energy density  of $\rho_E \sim 3 \times 10^{-19}\rm\,erg\
 cm^{-3}$, observed in cosmic rays and electromagnetic energy, ends up in neutrinos with a spectrum $E_\nu dN / dE_{\nu}  \sim E^{-\gamma}\rm\, cm^{-2}\, s^{-1}\, sr^{-1}$ that continues up to a maximum energy $E_{\rm max}$. The neutrino flux follows from the relation
%
$ \int E_\nu dN / dE_{\nu}  =  c \rho_E / 4\pi  $.
%
For $\gamma = 1$ and $E_{\rm max} = 10^8$\,GeV, the generic source of the highest energy cosmic rays produces a flux of $ {E_\nu}^2 dN / dE_{\nu}  \sim 5 \times 10^{-8}\rm\, GeV \,cm^{-2}\, s^{-1}\, sr^{-1} $.

There are several ways to modify this simple prediction:
\begin{itemize} 
\item The derivation fails to take into account the fact that  there are more cosmic rays in the universe producing neutrinos than observed at earth because of the GZK-effect and neglects evolution of the sources with redshift. This increases the neutrino flux by a factor $\sim$\,3, possibly more.
\item For proton-$\gamma$ interactions muon neutrinos receive only 1/4 of the energy of the charged pion in the decay chain $\pi^+\rightarrow \mu^+ +\nu_{\mu}\rightarrow e^+ +\nu_e +\bar{\nu}_{\mu} +\nu_{\mu}$ assuming that the energy is equally shared between the 4 leptons and taking into account that oscillations over cosmic distances distribute the neutrino energy equally among the 3 flavors.
\end{itemize}
The corrections approximately cancel.

Studying specific models of cosmic ray accelerators one finds that the energy supplied by the black hole to cosmic rays usually exceeds that transferred to pions, for instance by a factor 5 in the case of gamma ray bursts. We therefore estimate that the muon-neutrino flux associated with the sources of the highest energy cosmic rays is loosely confined to the range $ {E_\nu}^2 dN / dE_{\nu}= 1\sim 5 \times 10^{-8}\rm\, GeV \,cm^{-2}\, s^{-1}\, sr^{-1} $ yielding $10 \,{\sim}\, 50$ detected muon neutrinos per km$^2$ per year. This number depends weakly on $E_{\rm max}$ and the spectral slope~$\gamma$. The observed event rate is obtained by folding the predicted flux with the probability that the neutrino is actually detected in a high energy neutrino telescope; the latter is given by the ratio of the muon and neutrino interaction lengths in the detector medium, $\lambda_\mu / \lambda_\nu$\cite{PR}.

This flux has to be compared with the sensitivity of ${\sim}10^{-7}\rm\, GeV\ cm^{-2}\, s^{-1}\,sr^{-1}$ reached during the first 4 years of operation of the completed AMANDA detector in 2000--2003\cite{hill}. The analysis of the data has not been completed, but a preliminary limit of  $2.9 \times 10^{-7}\rm\,GeV\ cm^{-2}\,s^{-1}\,sr^{-1}$ has been obtained with a single year of data\cite{b10-diffuse}. On the other hand, after three years of operation IceCube will reach a diffuse flux limit of $E_{\nu}^2 dN / dE_{\nu} = 2\,{\sim}\, 7 \times 10^{-9}\rm\,GeV \,cm^{-2}\, s^{-1}\, sr^{-1}$. The exact value depends on the magnitude of the dominant high energy atmospheric neutrino background from the prompt decay of atmospheric charmed particles\cite{ice3}. The level of this background is difficult to anticipate. A cosmic flux at the ``Waxman-Bahcall" level will result in the observation of several hundred neutrinos in IceCube\cite{ice3}.

\section{Kilometer-Scale Detectors}

Arguing that a generic cosmic accelerator produces equal energies in cosmic rays, photons and neutrinos, we derived the  ``Waxman-Bahcall'' flux. A kilometer-scale detector is required to detect the roughly $10 {\sim} 50$ events per km$^2$ year. Model calculations assuming that active galaxies or gamma-ray bursts are the actual sources of cosmic rays yield similar, or even smaller event rates.

The case for kilometer-scale detectors also emerges from the consideration of  ``guaranteed'' cosmic fluxes. Neutrino fluxes are guaranteed when both the accelerator and the pion producing target material can be identified. Examples include:
\begin{itemize}
\item The extragalactic cosmic rays produce $0.1 \sim$ a few events per km$^2$ year in interactions with cosmic microwave photons. Furthermore, these cosmic rays are magnetically trapped in galaxy clusters and may produce additional neutrinos on the X-ray emitting gas in the cluster.
\item Galactic cosmic rays interact with hydrogen in the disk producing an observable neutrino flux in a kilometer-scale detector.
\item Air shower arrays have observed a ``directional'' flux of cosmic rays from the galactic plane, unlikely to be protons whose directions are scrambled in the magnetic field. The flux appears only in a narrow energy range from $1\,{\sim}\, 3$\,EeV, the energy where neutrons reach typical galactic kiloparsec distances within their lifetime of minutes. Both the directionality and the characteristic energy make a compelling case for electrically neutral neutron primaries. For every neutron reaching earth, a calculable number decays into electron antineutrinos before reaching us. Their flux should be observable in neutrino telescopes\cite{luis}: from the Cygnus region at the South Pole and from the galactic center for a Mediterranean detector.
\end{itemize}
In conclusion, observation of ``guaranteed'' sources also requires  kilometer-size neutrino detectors, preferably operated over many years.

Finally and most importantly, with recent observations\cite{hess} of the supernova remnant RX J1713.7-3946 using the H.E.S.S. atmospheric Cherenkov telescope array, gamma-ray astronomy may have pointed at a truly guaranteed source of cosmic neutrinos\cite{alvarezhalzen}.  The observations of TeV-gamma rays from the supernova remnant may have identified the first site where protons are accelerated to energies typical of the main component of the galactic cosmic rays\cite{hess}. Although the resolved image of the source (the first ever at TeV energies!) reveals TeV emission from the whole supernova remnant, it shows a clear increase of the flux in the directions of known molecular clouds. This naturally suggests the possibility that protons, shock accelerated in the supernova remnant, interact with the dense clouds to produce neutral pions that are the source of the observed increase of the TeV signal. Furthermore, the high statistics H.E.S.S. data for the flux are power-law behaved over a large range of energies without any signature of a cutoff characteristic of synchrotron or inverse-Compton sources. Other interpretations are not ruled out\cite{hiraga} but, fortunately, higher statistics data is forthcoming.

If future data confirms that a fraction of the TeV flux of RX J1713.7-3946 is of neutral pion origin, then the accompanying charged pions will produce a guaranteed neutrino flux of at least 20 muon-type neutrinos per kilometer-squared per year\cite{alvarezhalzen}. From a variety of such sources we can therefore expect event rates of cosmic neutrinos of galactic origin similar to those estimated for extragalactic neutrinos in the previous section. Supernovae associated with molecular clouds are a common feature of the OB associations that exist throughout the galactic plane. They have been suspected to be the sources of the galactic cosmic rays for some time.

It is important to realize that the relation between the neutrino and gamma flux is robust\cite{alvarezhalzen}. The $\nu_\mu + \bar\nu_\mu$ neutrino flux ($dN_\nu/dE_\nu$) produced by the decay of charged pions in the source can be derived from the observed gamma ray flux by imposing energy conservation:
\begin{equation}
\int_{E_{\gamma}^{\rm min}}^{E_{\gamma}^{\rm max}}
E_\gamma {dN_\gamma\over dE_\gamma} dE_\gamma = K
\int_{E_{\nu}^{\rm min}}^{E_{\nu}^{\rm max}} E_\nu {dN_\nu\over dE_\nu} dE_\nu
\label{conservation}
\end{equation}
where ${E_{\gamma}^{\rm min}}$ ($E_{\gamma}^{\rm max}$) is the minimum (maximum) energy of the photons that have a hadronic origin. ${E_{\nu}^{\rm min}}$ and ${E_{\nu}^{\rm max}}$ are the corresponding minimum and maximum energy of the neutrinos.
The factor $K$ depends on whether the $\pi^0$'s are of $pp$ or $p\gamma$ origin. Its value can be obtained from routine particle physics. In $pp$ interactions 1/3 of the proton energy goes into each pion flavor on average. In the pion-to-muon-to-electron decay chain 2 muon-neutrinos are produced with energy $E_\pi/4$ for every photon with energy $E_\pi/2$ (on average). Therefore the energy in neutrinos matches the energy in photons and $K=1$. This flux has to be reduced by a factor 2 because of oscillations. The estimate should be considered a lower limit because the photon flux to which the calculation is normalized, may be partially absorbed in the source or in the interstellar medium.

\section{Neutrino Telescopes: First ``Light''}

While it has been realized for many decades that the case for neutrino astronomy is compelling, the challenge has been to develop a reliable, expandable and affordable detector technology to build the kilometer-scale telescopes required to do the science. Conceptually, the technique is simple. In the case of a high-energy muon neutrino, for instance, the neutrino interacts with a hydrogen or oxygen nucleus in deep ocean water and produces a muon travelling in nearly the same direction as the neutrino. The Cerenkov light emitted along the muon's kilometer-long trajectory is detected by a lattice of photomultiplier tubes deployed on strings at depth shielded from radiation. The orientation of the Cerenkov cone reveals the roughly collinear muon and neutrino direction.

The AMANDA detector, using natural 1 mile-deep Antarctic ice as a Cerenkov detector, has operated for more than 4 years in its final configuration of 667 optical modules on 19 strings. The detector is in steady operation collecting roughly $7\,{\sim}\, 10$ neutrinos per day using fast on-line analysis software. The lower number will yield a background-free sample all the way to the horizon. AMANDA's performance has been calibrated by reconstructing muons produced by atmospheric muon neutrinos in the 50\,GeV to 500\,TeV energy range\cite{nature}.

Using the first 4 years of AMANDA\,II data, the AMANDA collaboration is performing a search for  the emission of muon neutrinos from spatially localized directions in the northern sky. Only the year 2000 data have been published~\cite{HS}. The skyplot for the 4 years of data is shown in Fig.\,3. A 90\% upper limit on the neutrino fluency of point sources is at the level of $6 \times 10^{-8}\rm\, GeV \,cm^{-2}\, s^{-1}$ or $10^{-10}\rm \,erg\ cm^{-2}\,s^{-1}$, averaged over declination. This corresponds to a flux of $6 \times 10^{-9}\rm\, cm^{-2}\, s^{-1}$ integrated above 10\,GeV assuming an $E^{-2}$ energy spectrum typical for shock acceleration of particles in high energy sources. In a search for 33 preselected sources, the most significant excess is from the Crab with 10 events where 5 are expected, not significant given the number of trials. IceCube is needed to make conclusive observations of sources. There are several ways to further improve the detector's reach, for instance by including time information on the sources provided by observations at other (photon) wavelengths.

\begin{figure}[t]
\centering\leavevmode
\includegraphics[width=5in]{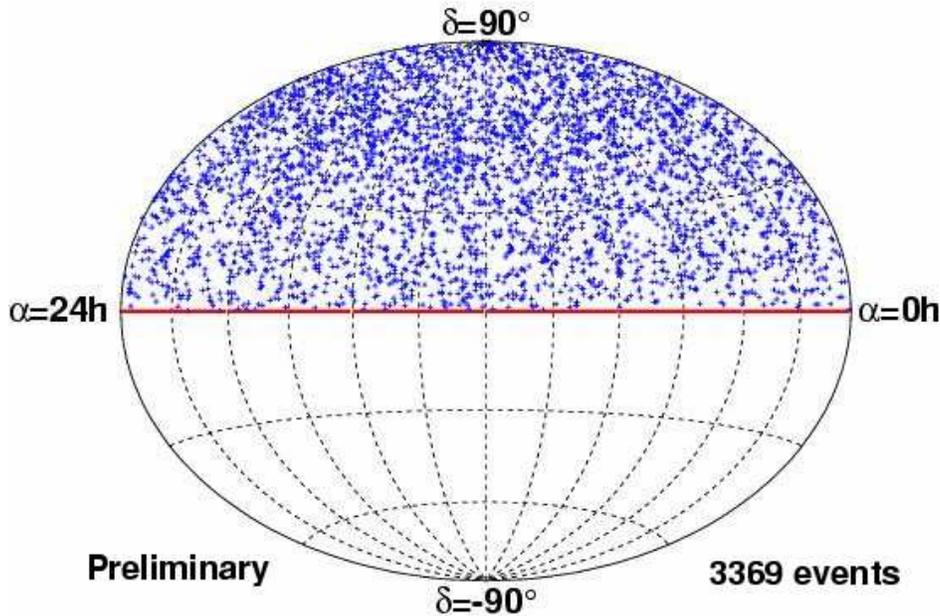}
\caption{Skymap showing declination and right ascension of neutrinos detected by the AMANDA\,II detector during four Antarctic winters of operation in 2000--2003.}
\end{figure}

The AMANDA\,II detector has reached a high-energy effective telescope area of 25,000$\,\sim\,	$40,000\,m$^2$, depending on declination. This represents an interesting milestone: known TeV gamma ray sources, such as the active galaxies Markarian 501 and 421, should be observed in neutrinos if the number of gamma rays and neutrinos emitted are roughly equal as expected from cosmic ray accelerators producing pions\cite{alvarezhalzen}. Therefore AMANDA must detect the observed TeV photon sources soon, or, its observations will exclude them as the sources of cosmic rays.

\section{Mediterranean Telescopes}

Below PeV energy, South Pole neutrino telescopes do not cover the Southern sky, which is obscured by the large flux of cosmic ray muons and neutrinos. This and the obvious need for more than one telescope --- accelerator experiments have clearly demonstrated the value of multiple detectors --- provide compelling arguments for deploying northern detectors. With the first observation of neutrinos by a detector in Lake Baikal with a telescope area of 2500\,m$^2$ for TeV muons\cite{baikal} and after extensive R\&D efforts by both the ANTARES\cite{antares} and NESTOR\cite{nestor} collaborations in the Mediterranean, there is optimism that the technological challenges to build neutrino telescopes in deep sea water have been met.  Both Mediterranean collaborations have demonstrated their capability to deploy and retrieve optical sensors, and have reconstructed down-going muons with optical modules deployed for R\&D tests.

The ANTARES neutrino telescope is under construction at a 2400\,m deep Mediterranean site off Toulon, France. It will consist of 12 strings, each equipped with 75 optical sensors mounted in 25 triplets. The detector performance has been fully simulated\cite{antares} with the following results: a sensitivity after one year to point sources of $0.4-5 \times 10^{-15}\rm\, cm^{-2}\, s^{-1}$ (note that this is the flux of secondary muons, not neutrinos) and to a diffuse flux of $0.9 \times 10^{-7}\rm\, GeV \,cm^{-2}\, s^{-1}$ above 50\,TeV. As usual, an $E^{-2}$ spectrum has been assumed for the signal. AMANDA\,II data have reached similar point source limits ($0.6 \times 10^{-15}\rm\, cm^{-2}\, s^{-1}\,sr^{-1}$) using 4 Antarctic winters, or about 1000 days, of data\cite{HS}). This value depends weakly on declination. Also the diffuse limits reached in the absence of a signal are comparable\cite{hill}. We have summarized the sensitivity of both experiments in Table~1, where they are also compared to the sensitivity of IceCube.

\begin{table}[h]
\caption{}
\tabcolsep1.25em
\def\arraystretch{1.5}
\begin{center}
\begin{tabular}{cccc}
\hline
& \bf IceCube& \bf AMANDA-II$^*$& \bf ANTARES\\
\hline
\bf \# of PMTs& 4800 / 10 inch& 600 / 8 inch& 900 / 10 inch\\
\hline
\parbox{6.5em}{{\bf Point source sensitivity} (muons/year)}& $6\times 10^{-17}\,\rm cm^{-2}\,\rm s^{-1}$&
\parbox{9em}{\centering $1.6\times 10^{-15}\,\rm cm^{-2}\, s^{-1}$ weakly dependent\\ on oscillations}&
\parbox{10em}{\centering 0.4--$5\times 10^{-15}\rm\, cm^{-2}\,s^{-1}$ depending\\ on declination}\\[.2in]
\hline
\parbox{6.5em}{{\bf diffuse limit$^\dagger$} (muons/year)}& 
\parbox{7.5em}{\centering 3--$12\times 10^{-9}\,\rm GeV\break cm^{-2}\,s^{-1}\,sr^{-1}$}&
\parbox{6em}{\centering $2\times 10^{-7}\rm\,GeV\break cm^{-2}\, s^{-1}\,sr^{-1}$}&
\parbox{6.5em}{\centering $0.8\times 10^{-7}\rm\,GeV\break cm^{-2}\,s^{-1}\,sr^{-1}$}\\[.1in]
\hline
\multicolumn{4}{l}{$^*$includes systematic errors}\\
\multicolumn{4}{l}{$^\dagger$depends on assumption for background from atmospheric neutrinos from charm}
\end{tabular}
\end{center}
\end{table}

Given that AMANDA and ANTARES operate at similar depths and have similar total photocathode area (AMANDA\,II is actually a factor of 2 smaller with 667 8-inch versus 900 10-inch photomultipliers for Antares), the above comparison provides us with a first glimpse at the complex question of the relative merits of water and ice as a Cherenkov medium. The conclusion seems to be that, despite differences in optics and in the background counting rates of the photomultipliers, the telescope sensitivity is approximately the same for equal photocathode area. The comparison is summarized in Table~1 where we have tabulated the sensitivity of AMANDA and Antares to points sources and to a diffuse flux of neutrinos. At this time, in the absence of a discovery, it is the sensitivity and not the area or angular resolution that represents the relevant quantity of merit. In the same context, the NEMO collaboration has done the interesting exercise of simulating the IceCube detector (augmented from 4800 to 5600 optical modules; see next section) in water rather than ice. One finds a slightly reduced sensitivity in water, probably not significant within errors and at no energy larger than 50\%\cite{emigneco}. Notice that in several years of operation a kilometer-scale detector like IceCube can improve the sensitivity of first-generation telescopes by two orders of magnitude.

\section{Kilometer-scale Neutrino Observatories}

The baseline design of kilometer-scale neutrino detectors maximizes sensitivity to $\nu_\mu$-induced muons with energy above hundreds of GeV, where the acceptance is enhanced by the increasing neutrino cross section and muon range and the Earth is still largely transparent to neutrinos. The mean-free path of a $\nu_\mu$ becomes smaller than the diameter of the earth above 70\,TeV --- above this energy neutrinos can only reach the detector from angles closer to the horizon. Good identification of other neutrino flavors becomes a priority, especially because $\nu_\tau$ are not absorbed by the earth. Good angular resolution is required to distinguish possible point sources from background, while energy resolution is needed to enhance the signal from astrophysical sources, which are expected to have flatter energy spectra than the background atmospheric neutrinos.

Overall, AMANDA represents a proof of concept for the kilometer-scale neutrino observatory, IceCube\cite{ice3}, now under construction. IceCube will consist of 80 kilometer-length strings, each instrumented with 60 10-inch photomultipliers spaced by 17~m. The deepest module is 2.4~km below the surface. The strings are
arranged at the apexes of equilateral triangles 125\,m on a side. The instrumented (not effective!) detector volume is a cubic kilometer. A surface air shower detector, IceTop, consisting of 160 Auger-style Cherenkov detectors deployed over 1\,km$^{2}$ above IceCube, augments the deep-ice component by providing a tool for calibration, background rejection and air-shower physics, as illustrated in Fig.~4.

\begin{figure}[h]
\centering\leavevmode
\includegraphics[width=6in]{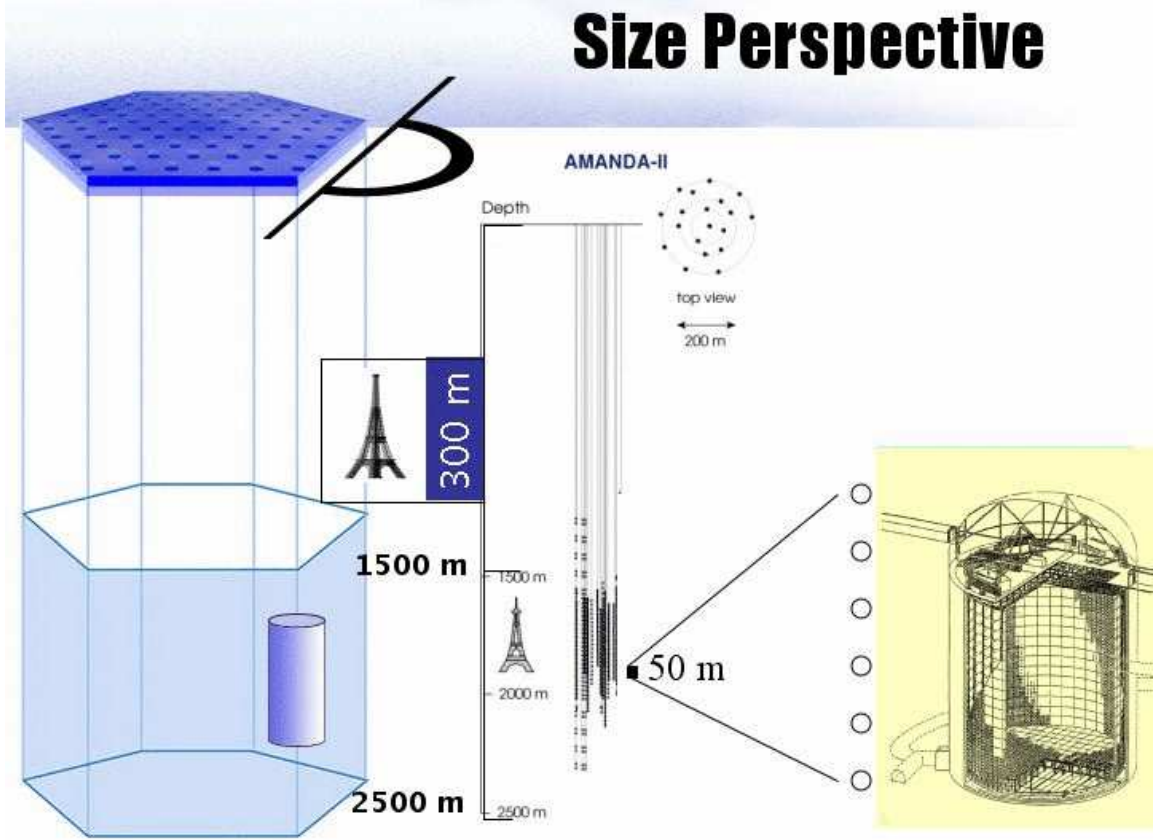}
\caption{Relative sizes of the IceCube, AMANDA, and Superkamiokande neutrino detectors. AMANDA will be operated as a lower threshold subsystem of IceCube. As the size of the detector grows, so does the threshold energy of neutrinos detected.}
\end{figure}

The transmission of analogue photomultiplier signals from the deep ice to the surface, used in AMANDA, has been abandoned. The photomultiplier signals will be captured and digitized inside the optical module.  The digitized signals are given a global time stamp with a precision of $<10$\,ns and transmitted to the surface.  The digital messages are sent to a string processor, a global event trigger and an event builder.  

Construction of the detector commences in the Austral summer
 of 2004/2005 and continues for 6 years, possibly less.  The growing detector will take data during construction, with each string coming online within days of deployment. The data streams of IceCube, and AMANDA\,II, embedded inside IceCube,  will be merged off-line using GPS timestamps.

IceCube will offer advantages over AMANDA\,II beyond its larger size: it will have a higher efficiency and superior angular resolution in reconstructing tracks, map showers from electron- and tau-neutrinos (events where both the production and decay of a $\tau$ produced by a $\nu_{\tau}$ can be identified) and, most
importantly, measure neutrino energy. Simulations, benchmarked by AMANDA data, indicate that the direction of muons can be determined with sub-degree accuracy and
their energy measured to better than 30\% in the logarithm of the energy. The direction of showers will be reconstructed to better than 10$^\circ$ above 10\,TeV and the response in energy is linear and better than 20\%. Energy resolution is critical because, once one establishes that the energy exceeds 1\,PeV, there is no atmospheric muon or neutrino background in a kilometer-square detector and full sky coverage of the telescope is achieved. The background counting rate of IceCube signals is expected to be less than 0.5\,kHz per optical sensor. In this low background environment, IceCube can detect the excess of MeV anti-$\nu_e$ events from a galactic supernova. 

NEMO, an INFN R\&D project in Italy, has been mapping Mediterranean sites and studying novel mechanical structures, data transfer systems as well as low power electronics, with the goal to deploy a next-generation detector similar to IceCube. A concept has been developed with 81 strings spaced by 140\,m. Each consists of 18 bars that are 20\,m long and spaced by 40\,m. A bar holds a pair of photomultipliers at each end, one looking down and one horizontally. As already mentioned, the simulated performance\cite{NEMO} is, not unexpectedly, similar to that of IceCube with a similar total photocathode area as the NEMO concept.

Recently, a wide array of projects have been initiated to detect neutrinos of the highest energies, typically above a threshold of 10 EeV, exploring other experimental signatures: horizontal air showers and acoustic or radio emission from neutrino-induced showers. Some of these experiments, such as the Radio Ice Cerenkov Experiment\cite{frichter} and an acoustic array in the Caribbean\cite{lehtinen}, have taken data; others are under construction, such as the Antarctic Impulsive Transient Antenna\cite{gorham}. The more ambitious EUSO/OWL project aims to detect the fluorescence of high energy cosmic rays and neutrinos from a detector attached to the International Space Stations.

\section*{Acknowledgments}
I thank my AMANDA/IceCube collaborators and Teresa Montaruli for discussions. This research was supported in part by the National Science Foundation under Grant No.~OPP-0236449, in part by the U.S.~Department of Energy under Grant No.~DE-FG02-95ER40896, and in part by the University of Wisconsin Research Committee with funds granted by the Wisconsin Alumni Research Foundation.

\end{document}